\documentclass{iau}

\usepackage{amsmath}
\usepackage{graphicx}
\usepackage{multirow}
\usepackage{natbib}

\begin{document}

\lefttitle{Gopalswamy {\it et al.}}
\righttitle{Cycle-25 halo CMEs}

\jnlPage{1}{7}
\jnlDoiYr{2024}
\doival{10.1017/xxxxx}
\volno{388}
\pubYr{2024}
\journaltitle{Solar and Stellar Coronal Mass Ejections}

\aopheadtitle{Proceedings of the IAU Symposium}
\editors{N. Gopalswamy,  O. Malandraki, A. Vidotto \&  W. Manchester, eds.}

\title{Implications of the abundance of halo coronal mass ejections for the strength of solar cycle 25}

\author{Nat Gopalswamy$^1$, Grzegorz Micha\l{}ek$^{2}$, Seiji Yashiro$^{3,1}$, Pertti M{\"a}kel{\"a}$^{3,1}$, Sachiko Akiyama$^{3,1}$ and Hong Xie$^{3,1}$}
\affiliation{{}$^1$NASA Goddard Space Flight Center, Greenbelt, MD, USA\\email: \email{nat.gopalswamy@nasa.gov}}
\affiliation{{}$^2$Jagiellonian University, Krak{\'o}w, Poland}
\affiliation{{}$^3$The Catholic University of America, Washington, DC, USA}

\begin{abstract}
We assess the relative strength of solar cycle (SC) 25 with respect to SCs 23 and 24 based on the abundance of halo coronal mass ejections (CMEs). We make use of the halo CME database (\url{https://cdaw.gsfc.nasa.gov/CME_list/halo/halo.html}) to compare the halo CME abundance during the first four years in each of SCs 23--25. The main result is that in several aspects such as the abundance, occurrence rate, source locations, and halo heights, halo CMEs are similar between SCs 24 and 25 but different from SC 23.  This result follows from the fact that weaker cycles have low heliospheric total pressure, whose backreaction on CMEs allows them to expand more and hence enhancing the chance of becoming a halo. The solar cycle variation of halo CME properties is consistent with the precursor-based cycle prediction methods that indicate SC 25 is similar to or only slightly stronger than SC 24.
\end{abstract}


\maketitle

\section{Introduction}

A coronal mass ejections (CME) is referred to a halo CME when its apparent width in the sky plane is 360$^\circ$. The occurrence rate of halo CMEs is correlated with the sunspot number (SSN), so it is likely that this correlation has a bearing on the prediction of cycle strengths. During the maximum phase of solar cycle (SC) 24 it was recognized that the halo CME abundance had a discordant behavior with SSN: the halo CME abundance normalized to SSN was higher in the weaker cycle 24 than in SC 23 \citep{2015ApJ...804L..23G,2022ApJ...936..122D}. This was explained to be a consequence of the anomalous expansion of CMEs due to the weak heliospheric state \citep{2014GeoRL..41.2673G}.  Comparing occurrence rates of halo CMEs during the rise phases of SCs 23--25, it was recently reported that the halo CME abundance is higher in SC 25 as well. Furthermore, the halo CME abundance indicates that the strength of SC 25 is intermediate between SCs 23 and 24, but closer to SC 24 \citep{2023ApJ...952L..13G}. Thus, the halo CME abundance provides clues to the relative strength of solar cycles.  Given the widely distributed predictions of the strength of SC 25, the halo CME abundance is able to discriminate among the predictions that vary from less than  half to almost twice the strength of SC 24 \citep{2021SoPh..296...54N}. In this paper, we extend the results of \citet{2023ApJ...952L..13G} to the maximum phase thereby including a bigger sample of CMEs over the first four years of cycles 23--25 to assess their relative strengths.

\section{Observations}

\begin{figure}[th]
  \centering
    \includegraphics[scale=.75]{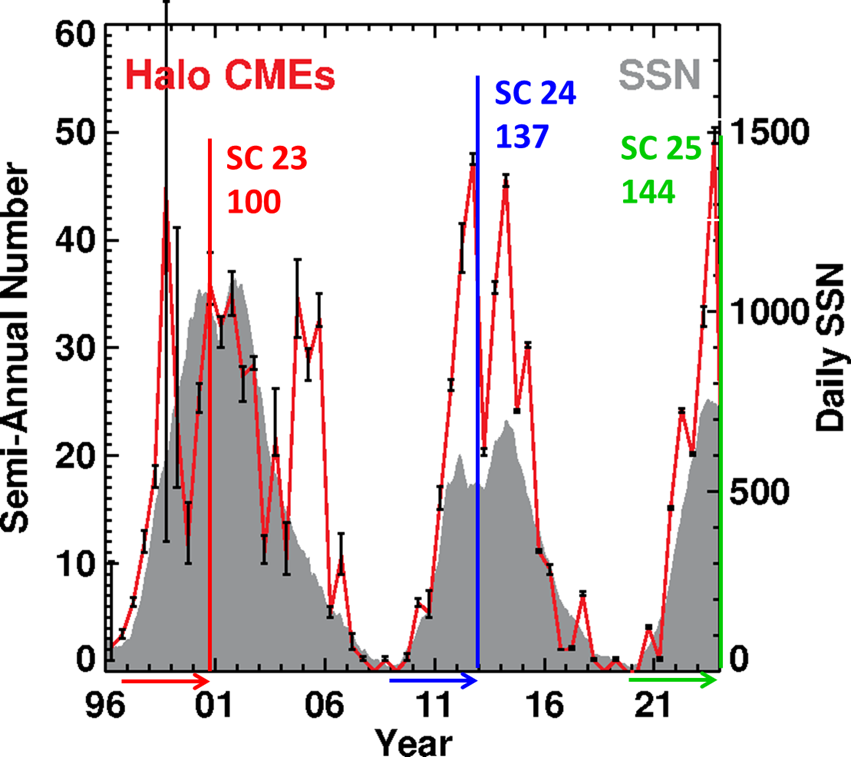}
    \caption{Semiannual variation of the number of halo CMEs. During the first 49 months (indicated by arrows), 100, 137, and 144 halos were observed in SCs 23, 24 and 25, respectively. SC 23: 1996 May 10 to 2000 June 9; SC 24: 2008 December 1 to 2012 December 31; SC 25: 2019 December 1 to 2023 December 31.  The error bars are based on the data gaps in the LASCO data. SSN (Source: WDC-SILSO, Royal Observatory of Belgium, Brussels) is shown for reference. }
    \label{fig1}
  \end{figure}

We use the halo CME catalog \citep[\url{https://cdaw.gsfc.nasa.gov/CME_list/halo/halo.html},][]{2010SunGe...5....7G} based on the observations from the Large Angle and Spectrometric Coronagraph \citep[LASCO;][]{1995SoPh..162..357B} on board the Solar and Heliospheric Observatory (SOHO). The catalog includes the solar sources of the halo CMEs (heliographic coordinates). We use the v2 SSN from the Sunspot Index and Long-term Solar Observations (SILSO, World Data Center, \url{https://sidc.be/SILSO/datafiles}). Figure~\ref{fig1} shows the semiannual number of halo CMEs as a function of time from 1996 to the end of 2023. We see that the number of halos during the first 49 months of SC 23 is only 100 compared to 137 in SC 24 and 144 in SC 25.  Thus, the halo CME abundance increased in SCs 24 and 25 although the SSN decreased  significantly.

\section{Analysis and Results}

\subsection{Halo CME abundance}

\begin{table}[b!]
\centering
\caption{Comparison of the number of halos in SCs 23--25}\label{table_1}
{\tablefont\begin{tabular}{@{\extracolsep{\fill}}lrrr}
\midrule
&SC 23&SC 24&SC 25\\
\midrule
Total \# Halos&100 (109)&137&144\\
\hspace{3mm}Change& & +26\% & +32\% \\
Avg SSN& 82.7 & 47.8 & 60.5 \\
\hspace{3mm}Change& & -42\% & -17\% \\
\#Halo/$<$SSN$>$&1.32&2.87&2.38\\
\hspace{3mm}Change& &+117\% & +80\% \\
\midrule
\end{tabular}}
\end{table}

Table~\ref{table_1} compares the number of halos during the first 49 months of each cycle and the relative changes with respect to cycle 23 as absolute values and relative to the average SSN. In cycle 23, there was a 4-month data gap. If we assume that halo CMEs occurred during the data gaps at the same rate as the average rate over the whole interval, the estimated number is 109 CMEs instead of 100. We use the corrected value (109) in comparisons. In terms of absolute numbers, we see that the number of halos increased by 26\% and 32\% in SCs 24 and 25, respectively. On the other hand, the average SSN declined by 42\% in SC 24 and 27\% in SC 25.  When we normalize the halo number to SSN, the relative numbers are 1.32, 2.87, and 2.38 per SSN. Clearly, the normalized halo CME abundance approximately doubled in SCs 24 and 25. Furthermore, the SC 24 abundance is the highest. The SC 25 abundance is much closer to that of SC 24.

  \begin{figure}[th]
  \centering
    \includegraphics[scale=.75]{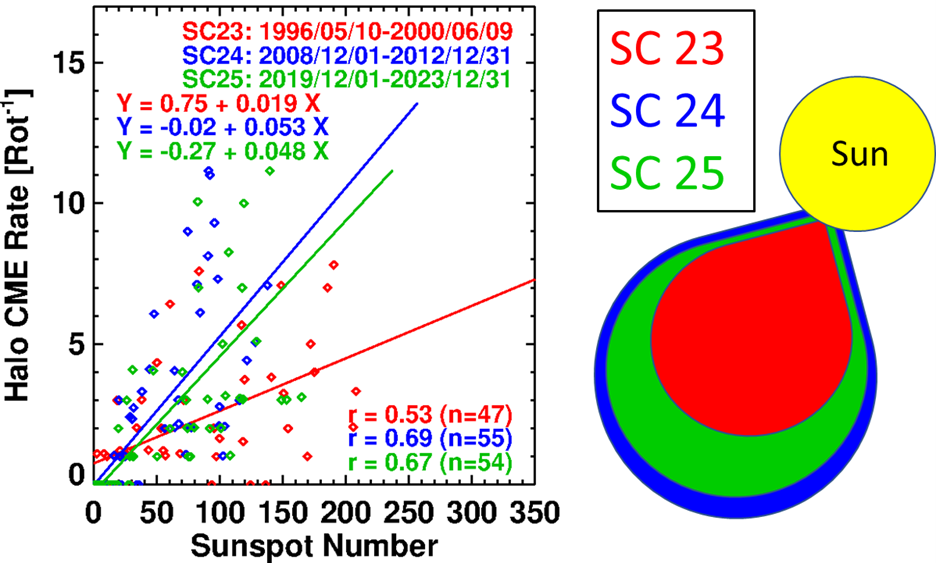}
    \caption{(left) Scatter plot between SSN and the occurrence rate of halo CMEs per Carrington rotation (CR) period. The number of data points during the first 49 months is not the same in the three cycles due to data gaps. The beginning and end dates of the compared intervals, the correlation coefficients and equations of the regression lines are shown on the plot. (right) Schematic showing the size CMEs in each cycle for a given CME speed.}
    \label{fig2}
  \end{figure}

\subsection{Correlation between SSN and halo CME occurrence rate}

Figure~\ref{fig2} shows the scatter plot between SSN and the number of halo CMEs in each Carrington rotation (CR) period. The SSN used in the plot is an average over CR periods.  We see that the halo CME rate is positively correlated with SSN in each of the three SCs. All correlations are significant because the Pearson's critical coefficients for 47, 55, and 54 samples are 0.372, 0.363, and 0.360, respectively for a p value of 1\% (the probability that the obtained correlation is by chance). The regression lines have similar slopes in SCs 24 and 25 but significantly higher than that in SC 23. These plots indicate that there are more halo CMEs for a given SSN in SCs 24 and 25, confirming the result shown in Table 1 for the whole interval.  The effect of the weak heliosphere on the CME size is shown schematically in Fig.~\ref{fig2}. The size of the CMEs is inversely related to the total pressure in the heliosphere. For a given speed, CMEs expand more in SCs 24 and 25 than in SC 23 resulting in the higher abundance of halo CMEs in SCs 24 and 25. 

  \begin{figure}[h]
  \centering
    \includegraphics[scale=0.75]{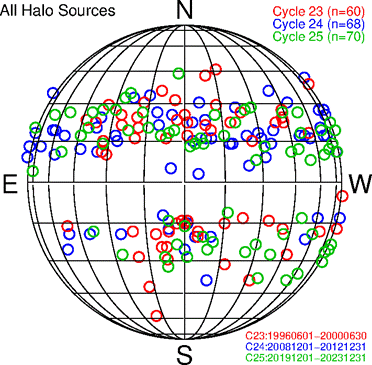}
    \caption{Solar source locations (heliographic coordinates) of halo CMEs during the first 49 months in SCs 23--25, color coded in red, blue, and green, respectively. Only frontside halos are included.}
    \label{fig3}
  \end{figure}

\subsection{Distribution of solar source locations}

The distribution of halo CME sources on the solar disk is shown in Fig.~\ref{fig3}.  The source locations are generally concentrated in the latitude range 15$^\circ$--30$^\circ$, which corresponds to the active region belt. Most halo CMEs originate from active regions, where large amount of free energy can be stored and released. Most of the source locations of SC 23 CMEs are concentrated near the disk center (within the longitude range of $\pm$45$^\circ$). On the other hand, the halos from SCs 24 and 25 originate from a broader range of longitudes. The mean absolute longitudes are 23$^\circ$.4 (SC 23), 37$^\circ$.0 (SC 24), and 36$^\circ$.2 (SC 25) with standard deviations of 20$^\circ$.7, 26$^\circ$.9, and 26$^\circ$.1, respectively. In SC 23, only one halo source is located beyond a central meridian distance (CMD) of 45$^\circ$ in the eastern hemisphere and 6 in the western hemisphere out of the 60 frontside halos (or $\sim$12\%). On the other hand, there are 25 (out of 68) and 24 (out of 70) halos beyond a CMD of 45$^\circ$ in SCs 24 and 25, respectively amounting to 37\% and 34\% of the frontside halos in these cycles.  Clearly, the large-CMD halos in SCs 24 and 25 outnumber those in SC 23 by a factor of $\sim$3. Once again, we see that the longitudinal distribution of halo CME sources in SCs 24 and 25 are close to each other, but much wider than that in SC 23. Halos originating beyond a CMD of 60$^\circ$ have a similar pattern: 4 out of 60 (or 7\%) in SC 23 compared to 17 out of 68 (or 25\%) in SC 24 and 15 out of 70 (or 21\%) in SC 25. We also note that the large-CMD halo sources in SC 25 are intermediate between SCs 23 and 24 but closer to SC 24.

\subsection{Halo heights of limb halo CMEs}

\begin{figure}[h]
  \centering
    \includegraphics[scale=.7]{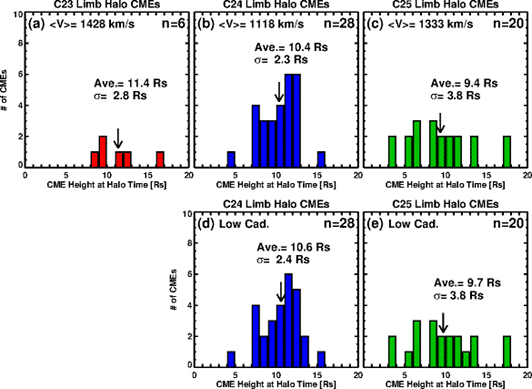}
    \caption{(top) Halo heights of limb CMEs in SCs 23--25 in panels a--c. The average of the distributions and their standard deviations are shown on the plots. The number of limb halos is higher than in Fig.~\ref{fig3} because we have included cases in which the sources are behind the limb but within 30$^\circ$. (bottom) Halo heights when the LASCO image cadence in SCs 24 and 25 reduced to match that in SC 23. Panels a--c display the average speeds of limb halos in three cycles. The speeds are obtained from \url{https://cdaw.gsfc.nasa.gov/CME_list/halo/halo.html}.}
    \label{fig4}
  \end{figure}

\citet{2020ApJ...897L...1G} introduced a new parameter known as "halo height" defined as the height of the leading edge of a limb CME when it appears as a halo for the first time in the LASCO FOV. In such limb halos, a CME originating close to one of the solar limbs ends up with features appearing on the opposite limb. Figure~\ref{fig4} shows the distributions of halo heights in the three cycles. The halo heights in SCs 24 and 25 are smaller than that in SC 23. Furthermore, the average speeds of the limb halos are lower in SCs 24 and 25. In other words, limb halos formed earlier in SCs 24 and 25 and at lower speeds. The limb halo sample size is small, so the results are not very firm. However, the trend is similar to what was seen between SCs 23 and 24 when the full-cycle data were used \citep{2020ApJ...897L...1G}.

\section{Discussion and Summary}

We investigated the properties of halo CMEs in SCs 23--25 that occurred during the first 49 months of each cycle, focusing on the abundance, occurrence rate, source locations, and halo heights. The results are consistent with the expected backreaction of the weaker heliospheric states in SCs 24 and 25 than that in SC 23. In all the aspects studied, we find that the halo CME properties are similar between SCs 24 and 25 and different from SC 23. Furthermore, the halo CME properties in SC 25 are intermediate between those in SCs 23 and 24, but closer to SC 24. These results strongly support the predictions of the strength of SC 25 based on precursor methods: the strength of SC 25 is similar to or slightly greater than that of SC 24 but much lower than that of SC 23. The trend that the space weather is milder in SC 24 than in SC 23 continues to SC 25 as well as indicated by the relatively smaller number of intense geomagnetic storms and high-energy particle events. The state of the heliosphere assessed from in-situ solar wind observations confirm the weaker state of the heliosphere in SC 25. These aspects of the three cycles will be reported elsewhere. The main results of this paper can be summarized as follows:

\begin{enumerate}
    \item The number of halo CMEs in SCs 23--24 is 109, 137, and 144 respectively during the first 49 months in each cycle. When normalized to SSN, the abundances become 1.32, 2.87, and 2.38 per SSN, respectively in SCs 23, 24, and 25.
    \item The number of halos binned by Carrington rotation periods are highly correlated with the corresponding CR-averaged SSN.  However, the slopes of the regression lines are steeper in SCs 24 and 25 than in SC 23, reflecting the higher abundance in the later cycles.
    \item The longitudinal distribution of halo CME sources is much wider in SCs 24 and 25 than in SC 23, e.g., the fraction of halos originating beyond a CMD of 60$^\circ$ is only 7\% in SC 23 compared to 25\% in SC 24 and 21\% in SC 25.
    \item The halo heights in SCs 24 and 25 are smaller than that in SC 23 but the average speeds of the limb halos are smaller than that in SC 23. In other words, limb halos form closer to the Sun at a smaller CME speed.
\end{enumerate}

We thank SOHO, STEREO, and SDO teams for making their data available online. The sunspot number is from WDC--SILSO, the Royal Observatory of Belgium, Brussels. SOHO is a joint project of ESA and NASA. STEREO is a mission in NASA’s Solar Terrestrial Probes program. Work supported by NASA’s LWS and STEREO programs. H.X. was partially supported by NSF grant AGS-2228967. P.M. was partially supported by NSF grant AGS-2043131.

\bibliography{halo}{}
\bibliographystyle{aasjournal}

\end{document}